\documentclass[prl,twocolumn,preprintnumbers,superscriptaddress,showpacs,amsmath,amssymb]{revtex4}
\usepackage{graphicx}
\usepackage{dcolumn}
\usepackage{bm}
\usepackage{color}
\usepackage[colorlinks, linkcolor=blue, urlcolor=blue, citecolor=blue, ]{hyperref}

\begin{document}

\title{Time-resolved observation of interatomic Coulombic decay induced by two-photon double excitation of Ne$_{2}$}

\author{T. Takanashi}\email{tsukasat@mail.tagen.tohoku.ac.jp}
\affiliation{Institute of Multidisciplinary Research for Advanced Materials, Tohoku University, 980-8577 Sendai, Japan}

\author{N. V. Golubev}
\affiliation{Theoretische Chemie, Universit\"{a}t Heidelberg, Im Neuenheimer Feld 229, 69120 Heidelberg, Germany}

\author{C. Callegari}
\affiliation{Elettra-Sincrotrone Trieste, Area Science Park, 34149 Basovizza, Trieste, Italy}

\author{H. Fukuzawa}\author{K. Motomura}\author{D. Iablonskyi}\author{Y. Kumagai}\author{\\S. Mondal}\author{T. Tachibana}
\affiliation{Institute of Multidisciplinary Research for Advanced Materials, Tohoku University, 980-8577 Sendai, Japan}

\author{K. Nagaya}\author{T. Nishiyama}
\affiliation{Department of Physics, Graduate School of Science, Kyoto University, 606-8502 Kyoto, Japan}

\author{K. Matsunami}
\affiliation{Department of Physics, Graduate School of Science, Kyoto University, 606-8502 Kyoto, Japan}

\author{P. Johnsson}
\affiliation{Department of Physics, Lund University, P.O. Box 118, 22100 Lund, Sweden}
	
\author{P. Piseri}
\affiliation{CIMAINA and Dipartimento di Fisica, Universit\`{a} di Milano, Via Celoria 16, I-20133 Milano, Italy}
	
\author{G. Sansone}\author{\\A. Dubrouil}\author{M. Reduzzi}\author{P. Carpeggiani}\author{C. Vozzi}
\author{M. Devetta}\author{M. Negro}\author{D. Faccial\`{a}}\author{F. Calegari}\author{A. Trabattoni}\author{ M. C. Castrovilli}
\affiliation{CNR-IFN, Piazza Leonardo da Vinci 32, 20133 Milano, Italy}

\author{Y. Ovcharenko}
\affiliation{Institut f\"{u}r Optik und Atomare Physik,  Technische Universit\"{a}t Berlin, Hardenbergstrasse 36, 10623 Berlin, Germany}

\author{M. Mudrich}\author{F. Stienkemeier}
\affiliation{Physikalisches Institut, Universit\"{a}t Freiburg, 79104 Freiburg, Germany}

\author{M. Coreno}
\affiliation{CNR-ISM,  Area Science Park, 34149 Basovizza, Trieste, Italy}

\author{\\M. Alagia}
\affiliation{CNR-IOM, Area Science Park, 34149 Basovizza, Trieste, Italy}

\author{B. Sch\"{u}tte}
\affiliation{Max-Born-Institut, Max-Born-Strasse 2 A, 12489 Berlin, Germany}

\author{N. Berrah}
\affiliation{Department of Physics, University of Connecticut, 2152 Hillside Road, Storrs, CT 06269, USA}

\author{O. Plekan}\author{P. Finetti}\author{C. Spezzani}\author{E. Ferrari}\author{\\E. Allaria}
\author{G. Penco}\author{C. Serpico}\author{G. De Ninno}\author{B. Diviacco}\author{S. Di Mitri}\author{L. Giannessi}
\affiliation{Elettra-Sincrotrone Trieste, Area Science Park, 34149 Basovizza, Trieste, Italy}

\author{\\G. Jabbari}
\affiliation{Theoretische Chemie, Universit\"{a}t Heidelberg, Im Neuenheimer Feld 229, 69120 Heidelberg, Germany}

\author{K. C. Prince}
\affiliation{CNR-IOM, Area Science Park, 34149 Basovizza, Trieste, Italy}
\affiliation{Elettra-Sincrotrone Trieste, Area Science Park, 34149 Basovizza, Trieste, Italy}

\author{L. S. Cederbaum}
\affiliation{Theoretische Chemie, Universit\"{a}t Heidelberg, Im Neuenheimer Feld 229, 69120 Heidelberg, Germany}

\author{\firstname{Ph.~V.} \surname{Demekhin}}
\affiliation{Institut f\"{u}r Physik und CINSaT, Universit\"{a}t Kassel, Heinrich-Plett-Str. 40, 34132 Kassel, Germany}

\author{A. I. Kuleff}
\affiliation{Theoretische Chemie, Universit\"{a}t Heidelberg, Im Neuenheimer Feld 229, 69120 Heidelberg, Germany}

\author{K. Ueda}
\email{ueda@tagen.tohoku.ac.jp}
\affiliation{Institute of Multidisciplinary Research for Advanced Materials, Tohoku University, 980-8577 Sendai, Japan}

\date{\today}

\begin{abstract}
The hitherto unexplored two-photon doubly-excited states [Ne$^{*}$($2p^{-1}3s$)]$_{2}$  were experimentally identified using the seeded, fully coherent, intense extreme ultraviolet free-electron laser FERMI. These states undergo ultrafast interatomic Coulombic decay (ICD) which predominantly  produces singly-ionized dimers. In order to obtain the rate of ICD, the resulting yield of Ne$_{2}^{+}$ ions was recorded as a function of delay between the XUV pump and UV probe laser pulses. The extracted lifetimes of the long-lived doubly-excited states, 390\,{\footnotesize({--130}/{+450})}~fs, and of the short-lived ones, less than 150~fs, are in good agreement with \emph{ab initio} quantum mechanical calculations.
\end{abstract}


\maketitle

About 20 years ago, it was predicted theoretically by Cederbaum \textit{et al.}  \cite{Cederbaum1997} that, if embedded in an environment, excited ionic species can decay non-radiatively by efficiently transferring their excess energy to this environment, which then releases the energy by emitting an electron. The process was termed interatomic/inter\-molecular Coulombic decay (ICD).  The first experimental observation of ICD was reported a few years later by Marburger \textit{et al.}~\cite{Marburger2003} who studied inner-valence ionization of Ne clusters by electron spectroscopy. Jahnke \textit{et al.}~\cite{Jahnke2004} then gave an unambiguous proof for the existence of ICD by an electron-ion-ion coincidence measurement in Ne dimers (Ne$_{2}$). Following these pioneering works, many experimental and theoretical studies have been reported in different systems and transitions (for recent reviews, see Refs.~\cite{Averbukh2011,Jahnke2015}). These extensive investigations demonstrated that ICD is relevant to various physical, chemical, and biological phenomena. It is worth noting that ICD was also observed in water~\cite{Mucke2010,Jahnke2010}, and its importance in biological systems surrounded by an aqueous environment was discussed in Refs.~\cite{Mucke2010,Jahnke2010,Stoychev2011}. The relevance of ICD to radiation therapy is also under discussion \cite{Gokhberg2014,Trinter2014}.

Not only the spectroscopic aspects, but also the dynamic aspects of ICD have been studied extensively over the years, however, mostly by theory~\cite{Santra2000,Santra2001,Santra2002,Averbukh2006,Kuleff2007}. It has been shown, for example, that the ICD rates depend on the distance to the neighboring species, as well as on the number of neighbors, making the dynamics during an ICD process rather involved. This, together with the extreme efficiency of these processes (typically ICD  takes place on a femtosecond time scale), explains why there are only a few reports of time-resolved observations of ICD~\cite{Trinter2013,Schnorr2013}, besides the indirect extraction of the ICD rates from the spectral profile measurements~\cite{Ohrwall2004,Ouchi2011}. With the advent of extreme ultraviolet (XUV) free-electron lasers (FELs), direct time-resolved investigations of ICD became possible and some promising approaches have been suggested~\cite{Demekhin2011a,Sansone2012,Dubrouil2015}. XUV FELs provide unprecedented high photon flux with  extremely short pulses of less than 100~fs \cite{Allaria2012,Allaria2013a}. Thus, a pump-probe measurement with well synchronized laser pulses may provide direct access to the time-evolution of ICD. In this letter, we perform a time-resolved study of ICD which makes use of the XUV pump\,--\,UV probe technique.

Stimulated by the  developments of XUV FELs, a new class of ICD processes in the multiply-excited clusters was recently predicted theoretically \cite{Kuleff2010}. In these processes, transfer of the de-excitation energy from one of the excited atoms in a cluster results in  the ionization of another excited atom. This mechanism  plays a central role in the creation of ions when clusters are exposed to moderate intensity laser pulses of photon energies insufficient for a single-photon ionization. Recently, such a process was indirectly observed in helium droplets \cite{LaForge2014} and clusters \cite{Ovcharenko14}. At the same time, \emph{ab initio} dynamical calculations reported in Ref.~\cite{Demekhin2013} propose an efficient scheme for production of the doubly-excited Ne dimers by a single intense XUV pulse and predict respective ICD rates and electron spectra. Here, we report the first direct observation of those doubly-excited states in Ne$_2$ and measure their ICD lifetime.

The presently studied process consists of the two-photon double excitation of Ne dimers by an intense XUV pulse
\begin{equation}
\label{step1}
\mathrm{Ne}_{2} + 2\,\hbar\,\omega_{_{XUV}} \rightarrow [\mathrm{Ne}^{*}(2p^{-1}3s)]_{2},
\end{equation}
which is then followed by the ICD transition
\begin{equation}
\label{step2}
[\mathrm{Ne}^{*}(2p^{-1}3s)]_{2} \rightarrow \mathrm{Ne}^{+}(2p^{-1})\mathrm{Ne} + e_{_{ICD}}^{-}.
\end{equation}
The \emph{ab initio} potential energy curves of the relevant electronic states of Ne$_2$ are collected in Fig.~1 of Ref.~\cite{Demekhin2013}. From this figure one can recognize that almost all ICD final states $\mathrm{Ne}^{+}(2p^{-1})\mathrm{Ne}$  except of a single repulsive state are bound. As a consequence, the ICD process (\ref{step2}) produces predominantly stable singly-charged dimers  Ne$_2^+$ as demonstrated in Ref.~\cite{Demekhin2013}. In the present experiment, we additionally apply delayed UV probe pulse, whose  photon energy is just sufficient to ionize a 3s-electron of one of the excited Ne atoms in the dimer
\begin{multline}
\label{step3}
[\mathrm{Ne}^{*}(2p^{-1}3s)]_{2} + \hbar\,\omega_{_{UV}}   \rightarrow \\ \mathrm{Ne}^{+}(2p^{-1}) + \mathrm{Ne}^{*}(2p^{-1}3s) + e^{-}.
\end{multline}
The potential energy curves of  the excited ionic states  $\mathrm{Ne}^{+}(2p^{-1})\mathrm{Ne}^{*}(2p^{-1}3s)$  are computed in the present work using the method described in Refs.~\cite{Trofimov1995,Averbukh05,Kopelke11}. They are weakly-bound with a very shallow  minimum at about 7~\AA. Around the equilibrium internuclear distance of 3.1~\AA, where these states are expected to be populated by the UV probe pulse, the
curves exhibit a steep slope leading to dissociation of this population. Therefore, after interaction with the UV pulse the dimer will dissociate as indicated in Eq.~(\ref{step3}), producing $\mathrm{Ne}^{+}$ and  $\mathrm{Ne}^{*}$ fragments with a kinetic energy release of about 2 eV.

The present experiment was performed at the Low Density Matter (LDM) beam line~\cite{Lyamayev2013,Svetina2015} at FERMI \cite{Allaria2013b,Allaria2015,Callegari2016}. The circularly polarized XUV FEL beam was focused by a Kirkpatrick-Baez (KB) mirror system to a focal size of 30~$\mu$m FWHM. The pulses had an average energy of 32~$\mu$J and a duration  between 60 and 80~fs FWHM. The resulting peak intensity was estimated to be about 6.5$\times$10$^{13}$~W/cm$^2$. The repetition rate of the XUV pulse was set to 10~Hz. The Ne dimers were produced by adiabatic expansion of the Ne gas through a 100~$\mu$m nozzle and a pulsed valve with an opening time of 25~$\mu$s, which was synchronized with the XUV pulse. The stagnation pressure was set to 0.8~MPa and the nozzle temperature to 190~K. The Ne$^{+}$ and Ne$_{2}^{+}$ ions produced in steps (\ref{step2}) and (\ref{step3})  were detected by a time-of-flight (TOF) mass spectrometer. The $^{20}$Ne$^+_2$ signal was used to measure the yield of ionized dimers, whereas, in order to avoid saturation effects in the  $^{20}$Ne$^{+}$ signal, the  $^{22}$Ne$^{+}$ signal was used to measure the  yield of  ionized monomers.

\begin{figure}
\includegraphics[scale=0.3]{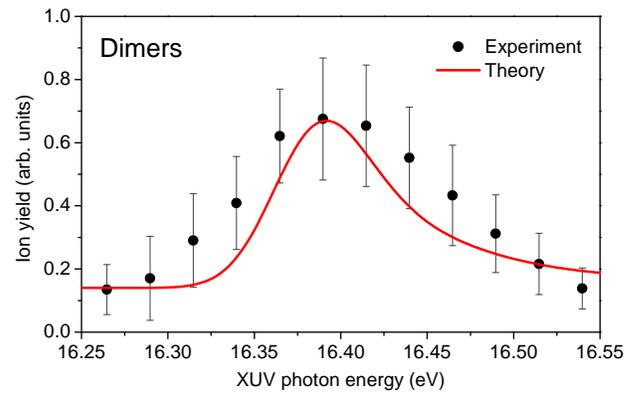}
\caption{The total yield of Ne$_{2}^{+}$ ions as a function of the XUV FEL photon energy measured (circles) without UV probe pulse and calculated (solid line) as described in Ref.~\cite{Demekhin2013}. The theoretical curve is shifted vertically by a constant to account for the background
in the experimental signal.}\label{fig:energy}
\end{figure}

To locate the theoretically predicted two-photon doubly-excited resonance, the XUV FEL photon energy was scanned within the range from 16.265 to 16.540~eV by 0.025~eV steps, seeking for the two-photon transition (\ref{step1}). The spectrum of each XUV pulse was recorded on a shot-by-shot basis and used to determine the central photon energy of the pulse by a Gaussian fit. The yield of  Ne$_{2}^{+}$ ions measured as a function of central photon energy is depicted in Fig.~\ref{fig:energy}. This yield exhibits a clear maximum at the photon energy of 16.39~eV. According to the theoretical predictions (see Fig.~2 in Ref.~\cite{Demekhin2013}), exactly this photon energy should be resonant for the two-photon transition into the [Ne$^{*}$($2p^{-1}3s$)]$_{2}$  doubly-excited states, which then produce stable Ne$_2^+$ ions by ICD (\ref{step2}). The total yield of singly-ionized dimers simulated for the present pulse parameters is also depicted in Fig.~\ref{fig:energy} (see Ref.~\cite{Demekhin2013} for details of calculations). The figure shows good agreement between the measured and computed Ne$_{2}^{+}$ yields  including the positions of the maxima and the asymmetry in their shapes, which are both skewed on the high-energy side.

Knowing that the used XUV pulses produce [Ne$^{*}$($2p^{-1}3s$)]$_{2}$  doubly-excited states, we fixed the XUV photon energy at 16.39~eV and performed time-resolved measurements using a delayed UV laser pulse as a probe. In these pump--probe measurements, the energy of the XUV pulse was set to 16~$\mu$J on average corresponding to a peak intensity of about 3.3$\times$10$^{13}$~W/cm$^2$. The photon energy of the probe UV laser pulse was 4.75~eV, its duration to 200~fs, and the average pulse energy was about 35~$\mu$J. The UV pulse was focused to the reaction point with a 80~$\mu$m focal size. The estimated average peak intensity was  about 8$\times$10$^{12}$~W/cm$^2$. The difference between the arrival times of the XUV and UV pulses, i.e., the time delay, was varied using an optical delay-line installed within the  path of the UV pulse.

\begin{figure}
\includegraphics[scale=0.4]{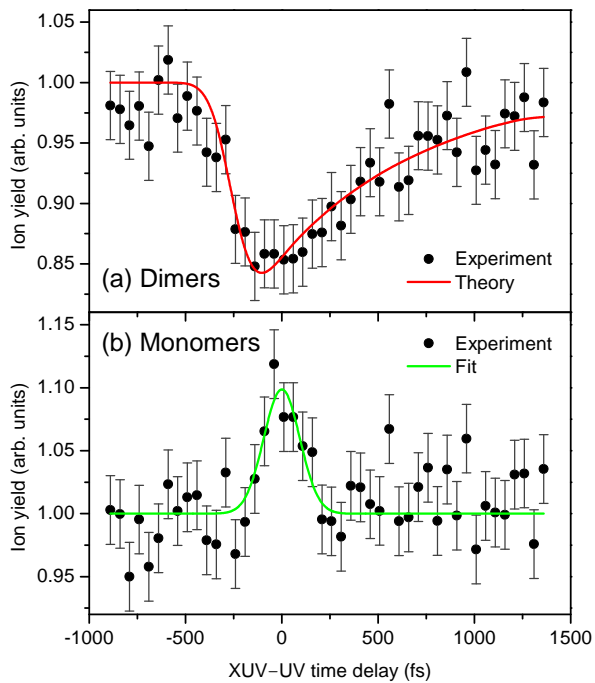}
\caption{\emph{Panel (a):} Total yield of Ne$_{2}^{+}$ ions measured (circles) as a function of the time delay between the XUV and UV pulses. The solid line is the result of model calculations described in the text. \emph{Panel (b):}  Total yield of Ne$^{+}$ ions measured (circles) as a function of the time delay. The ion yields of Ne dimers and monomers are collected over several tens of scans. In each scans the respective yields are normalized by the UV-off data in each delay point, in order to compensate for the effect of sampling dispersion. The zero point of the delay is calibrated via  Gaussian fit of the delay-dependent Ne$^{+}$ ion yield (shown as solid curve in  panel (b) to guide the eye).}\label{fig:time}
\end{figure}

Figure~\ref{fig:time} displays the presently measured yields of the Ne$_{2}^{+}$ and Ne$^{+}$ ions (circles with error bars) as functions of the time delay between the pump and probe pulses. The yield of Ne$_{2}^{+}$ ions (Fig.~\ref{fig:time}a) exhibits a clear dip around  zero time delay. The role of the UV pulse is to ionize the doubly-excited dimers, quenching thereby ICD and producing the dissociative $\mathrm{Ne}^{+}(2p^{-1})+\mathrm{Ne}^{*}(2p^{-1}3s)$ states in Eq.~(\ref{step3}). As a consequence, the yield of Ne$_{2}^{+}$ ions produced via ICD in Eq.~(\ref{step2}) decreases. Another consequence is that the yield of Ne$^{+}$ ions  increases as clearly seen in Fig.~\ref{fig:time}b around zero time delay. For each scan, this cross-correlation peak was fitted by a Gaussian function  in order to calibrate the zero point of the delay-axis.  The fitting procedure allows to determine the time delay with an accuracy of about 20~fs for the presently used XUV and UV pulses of  70 and 200~fs duration, respectively.

\begin{figure}
\includegraphics[scale=0.3]{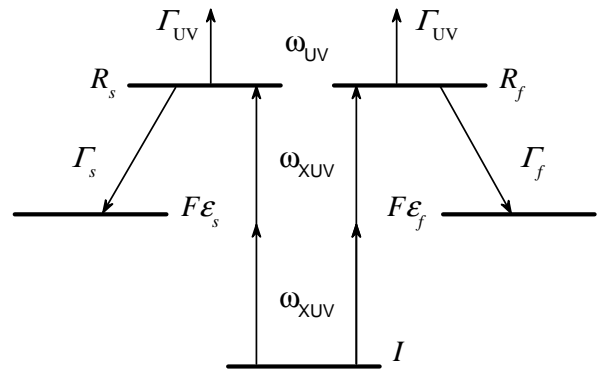}
\caption{Theoretical model used to simulate the presently measured delay-dependent yield of  Ne$_{2}^{+}$ ions and to extract the corresponding ICD transition rates. It includes two  doubly-excited states $\vert R_s \rangle$ and $\vert R_f \rangle$, each accessible by two XUV photons from the ground neutral state $\vert I \rangle$ and decaying by ICD into the final ionic state $\vert F\rangle$ with the emission of an ICD electron $\varepsilon_s$ or $\varepsilon_f$. The decay of  $\vert R_s \rangle$  state is \emph{slow} and that of $\vert R_f \rangle$  it is \emph{fast}, and the respective decay rates are $\Gamma_s$ and $\Gamma_f$. Alternatively, the UV pulse may ionize these excited states with the rate $\Gamma_{_{UV}}$, quenching thereby  ICD.}\label{fig:scheme}
\end{figure}

After a steep decrease at time delays between --400 and --100~fs, the measured Ne$_{2}^{+}$ yield in Fig.~\ref{fig:time}a increases over a large range of positive delays of more than 1500~fs, with a slope reflecting the lifetime of ICD. In the present work, we used this delay-dependent Ne$_{2}^{+}$ ion yield  to assess the ICD lifetimes (decay rates) as discussed below. The doubly-excited  [Ne$^{*}$($2p^{-1}3s$)]$_{2}$ dimers possess two types of \emph{gerade} states which are accessible from the ground Ne$_2(^1\Sigma_g^+)$ state by the absorption of two photons (see Fig.~1 of Ref.~\cite{Demekhin2013}). The two states of $^{1}\Sigma_{g}^{+}$ symmetry have relatively large total ICD rates (4.9~meV at 3.1~\AA) and are short-lived ($\tau_{_{ICD}}\sim130$~fs). The computed total ICD rates (2.1~meV at 3.1~\AA) of the remaining $^{1}\Pi_g$ and $^1\Delta_{g}$  states are smaller and these states are long-lived  ($\tau_{_{ICD}}\sim310$~fs). It is therefore important to distinguish  short- and long-lived doubly-excited states when analyzing the present time-resolved measurements.

To extract ICD rates from the experimental data, we have performed  dynamical simulations which reflect the essential physical mechanisms involved in the processes (\ref{step1}--\ref{step3}). In our theoretical model we consider a system of levels (see Fig.~\ref{fig:scheme}) which includes: the ground neutral electronic state $\vert I \rangle$, the short-lived $\vert R_f \rangle$ and the long-lived $\vert R_s \rangle$ doubly-excited states (designated by subscripts $f$  and $s$  for fast  and slow decay, respectively), and the two respective  ionic continua  $\vert F\varepsilon_f\rangle $ and $\vert F \varepsilon_s\rangle$. In the calculations, we used  laser pulses of the following form:
\begin{subequations}
\begin{equation}\label{XUVp}
\mathcal{E}_{_{XUV}}(t)= \mathcal{E}^{_{XUV}}_0 g_{_{XUV}}(t,\tau_{_{XUV}})\,\cos(\omega_{_{XUV}}t),
\end{equation}
\vspace{-0.75cm}
\begin{equation}\label{UVp}
\mathcal{E}_{_{UV}}(t)= \mathcal{E}^{_{UV}}_0 g_{_{UV}}(t,\tau_{_{UV}},\Delta t)\,\cos(\omega_{_{UV}}t).
\end{equation}
\end{subequations}
Here, $\mathcal{E}_0$ stands for the peak amplitude of the pulse, envelope  $g(t,\tau)$  of which has a Gaussian shape of duration  $\tau$. The time delay between the pump and probe pulses is denoted by $\Delta t$.

Within the rotating wave and local approximations, the  time evolution of the amplitudes of the populations of the initial and excited states, $a_I(t)$, $a_{R_s}(t)$, and $a_{R_f}(t)$, are given by the following system of coupled differential equations (see, e.g., Ref.~\cite{Demekhin2011b,Demekhin2012a,Demekhin2012b} for details of derivation)
\begin{subequations}
\label{aIR}
\vspace{-0.5cm}
\begin{equation}\label{aI}
i\dot{a}_I = \sqrt{2} D \left(\frac{\mathcal{E}^{_{XUV}}_0g_{_{XUV}}}{2}\right)^2 a_{R_s}+D \left(\frac{\mathcal{E}^{_{XUV}}_0g_{_{XUV}}}{2}\right)^2 a_{R_f},
\end{equation}
\vspace{-0.5cm}
\begin{equation}\label{ars}
i\dot{a}_{R_s} =\sqrt{2} D \left(\frac{\mathcal{E}^{_{XUV}}_0g_{_{XUV}}}{2}\right)^2 a_{I}-i\left(\frac{\Gamma_s}{2}+\frac{\Gamma_{_{UV}}g^2_{_{UV}}}{2}\right)a_{R_s},
\end{equation}
\vspace{-0.5cm}
\begin{equation}\label{arf}
i\dot{a}_{R_f} = D \left(\frac{\mathcal{E}^{_{XUV}}_0g_{_{XUV}}}{2}\right)^2 a_{I}-i\left(\frac{\Gamma_f}{2}+\frac{\Gamma_{_{UV}}g^2_{_{UV}}}{2}\right)a_{R_f}.
\end{equation}
\end{subequations}
In these equations, the energy of state $\vert I \rangle$ was set to zero, and the XUV photon energy is resonant for the two-photon excitations $\vert I \rangle \to \vert R_{s/f} \rangle$. The matrix element for this two-photon transition $D$ is a parameter. In order to avoid saturation in the excitation step, $D$  was chosen such that the pump XUV pulse promotes 10\% of the ground state population into the excited states. To  account for the double degeneracy of $\Pi$ and $\Delta$ vs. $\Sigma^+$ states, we introduce $\sqrt{2}$ for the transition into the long-lived state $\vert R_s\rangle $.

The ICD process (\ref{step2}) enters Eqs.~(\ref{ars}) and (\ref{arf}) as a leakage term $-\frac{i}{2}\Gamma_{s/f}$ of the populations of the $\vert R_{s/f} \rangle$ states by the corresponding decay rates. The time-dependent leakage of the populations of these excited states due to their ionization by the probe UV pulse Eq.~(\ref{step3}) is described in the equations by the imaginary term $-\frac{i}{2}\Gamma_{_{UV}}g^2_{_{UV}}$ \cite{Demekhin2011b}. Here, the total ionization  rate, $\Gamma_{_{UV}}=\sigma \, \Phi_{_{UV}}$, is a product of the photoionization  cross section  $\sigma$ and the flux $\Phi_{_{UV}}$ of the UV pulse.  The time evolution of the amplitudes of the population of the ICD final continuum states $\vert F\varepsilon_s\rangle $ and $\vert F \varepsilon_f\rangle$ are given by
\begin{subequations}
\label{aFF}
\begin{equation}\label{afs}
i\dot{a}_{F\varepsilon_s} = \sqrt{\frac{\Gamma_s}{2\pi}}a_{R_s} + (E_F+\varepsilon_s-2\,\omega_{_{XUV}})a_{F\varepsilon_s},
\end{equation}
\vspace{-0.5cm}
\begin{equation}\label{aff}
i\dot{a}_{F\varepsilon_f} = \sqrt{\frac{\Gamma_f}{2\pi}}a_{R_f} +  (E_F+\varepsilon_f-2\omega_{_{XUV}})a_{F\varepsilon_f}.
\end{equation}
\end{subequations}
Calculations via Eqs.~(\ref{aIR}) and (\ref{aFF}) need to be performed at each time delay  $\Delta t$ between the pump and probe pulses for all energies of the ICD electrons $\varepsilon_s$ and  $\varepsilon_f$. The total yield of Ne$_2^+$ ions can finally be obtained as a sum of the populations of the final ionic states integrated over all electron energies after both pulses have expired
\begin{equation}\label{yield}
S(\Delta t) = \lim_{t\to\infty} \sum_{\varepsilon=\varepsilon_s,\varepsilon_f}\int \vert a_{F\varepsilon}(t,\Delta t)\vert^2 d\varepsilon.
\end{equation}

In the calculations, we performed a fitting of the theoretical delay-dependent yield of Ne$_2^+$ ions (\ref{yield}) to the experimental data set in Fig.~\ref{fig:time}a. The present optimization procedure is based on the iterative Levenberg-Marquardt algorithm  for multivariate functionals \cite{Levenberg1944,Marquardt1963}. The set of optimization parameters includes decay rates of the slow $\Gamma_s$ and fast $\Gamma_f$ components of ICD. We also varied the impact of the probe UV pulse through $\Gamma_{_{UV}}$. This quantity is responsible for the depletion depth in the  Ne$_2^+$ yield (Fig.~\ref{fig:time}a) and it is determined by the photoionization cross section $\sigma$, which was the third variational parameter. Calculations were performed for the experimental parameters of XUV and UV pulses, i.e.,  $\omega_{_{XUV}}=16.39$~eV, $\tau_{_{XUV}}=70$~fs, $I_0^{_{XUV}}=3.3\times10^{13}$~W/cm$^2$ and $\omega_{_{UV}}=4.75$~eV, $\tau_{_{UV}}=200$~fs, $I_0^{_{UV}}=8\times10^{12}$~W/cm$^2$. The optimized delay-dependent ion yield  (\ref{yield}), computed as described above, is shown in Fig.~\ref{fig:time}a by the solid curve.

One can see from  Fig.~\ref{fig:time}a,  that the present model dynamical calculations reproduce the temporal profile of the experimental Ne$_2^+$ yield very well. The success of the present few-level model (Fig.~\ref{fig:scheme}), which neglects the nuclear wave packet dynamics accompanying process  (\ref{step1}--\ref{step3}), can be rationalized as follows. The potential energy curves of the involved doubly-excited states [Ne$^{*}$($2p^{-1}3s$)]$_{2}$ have their minima around 3.05--3.15\AA, which is very close to the equilibrium internuclear distance 3.1~\AA\, of the ground electronic state Ne$_2(^1\Sigma_g^+)$ (see Fig.~1 in Ref.~\cite{Demekhin2013}). As a consequence, the excitation and decay transitions in the Ne dimer are essentially vertical, taking place mainly around 3.1~\AA, and the nuclear dynamics awakes only in the ICD final states \cite{Demekhin2013}.

The present fitting procedure yields  $\Gamma_s=1.67\pm0.89$~meV  for the slow ICD rate corresponding to an ICD lifetime of 390\,{\footnotesize({--130}/{+450})}~fs. This value is in good agreement with the theoretical prediction of 2.1~meV at 3.1~\AA\, for the long-lived $^{1}\Pi_g$ and $^1\Delta_{g}$ doubly-excited states \cite{Demekhin2013}. The excitation and decay of the short-lived state $\vert R_s\rangle $ influences the computed ion yield (\ref{yield}) only slightly: This state decays so fast that for positive delays the probe UV pulse is too late to ionize it. Therefore, in the present work we were only able to estimate the lower limit of the fast ICD rate  $\Gamma_f$. The fitting procedures indicates that it should be larger than 4.5~meV, i.e., the corresponding lifetime should be shorter than 150~fs. This estimate agrees very well with the value of 4.9~meV at 3.1~\AA, reported in Ref.~\cite{Demekhin2013} for the two short-lived $^{1}\Sigma_{g}^{+}$ doubly-excited states.

In conclusion, using the fully coherent XUV pulses from the seeded free-electron laser FERMI, we have unambiguously identified the two-photon doubly-excited [Ne$^{*}$($2p^{-1}3s$)]$_{2}$ states of Ne dimer. The presently realized pump-probe scheme enabled direct access to the time evolution of ICD of these states. In particular, measuring the yield of Ne$_2^+$ ions as  a function of time delay between the pump XUV and probe UV pulses, we were able to determine the corresponding  ICD transition rates  (lifetimes). The experimental lifetimes of the short- and long-lived doubly-excited states  [Ne$^{*}$($2p^{-1}3s$)]$_{2}$, obtained with the help of the model dynamical calculations, are in  good agreement with the previously reported \textit{ab initio} values from Ref.~\cite{Demekhin2013}. We believe, that the presently realized scheme involving the XUV FEL pump and femtosecond UV probe pulses may become a powerful tool for future real-time dynamical investigations of ultrafast relaxation processes of excited systems and mechanisms of energy and charge transfer in media.

\begin{acknowledgements}
This work was supported
by the X-ray Free Electron Laser Priority Strategy Program of the Ministry of Education, Culture, Sports, Science, and Technology of Japan (MEXT);
by the Grant-in-Aid for the Global COE Program `the Next Generation of Physics, Spun from Universality and Emergence' from the MEXT;
by the Grants-in-Aid (20310055, 21244062) from the Japan Society for the Promotion of Science (JSPS);
by the Swedish Research Council and the Swedish Foundation for Strategic Research;
by the ERC Starting Research Grant UDYNI No. 307964;
by the Italian Ministry of Research and Education (ELI project - ESFRI Roadmap);
by the State of Hesse LOEWE focus-project ELCH;
by the European Research Council (ERC) Advanced Investigator Grant No. 692657; and
by the Deutsche Forschungsgemeinschaft (Forschergruppe FOR~1789 and SPP~1840/1 QUTIF).
\end{acknowledgements}

\end{document}